\documentclass[11pt]{article}
\usepackage{setspace}
\usepackage{amssymb}
\usepackage{amsmath}
\usepackage{amsfonts}
\usepackage{slashed}

\def\be{\begin{equation}}

\def\ee{\end{equation}}

\def\bea{\begin{eqnarray}}

\def\eea{\end{eqnarray}}

\makeatletter
\def\blfootnote{\xdef\@thefnmark{}\@footnotetext}
\makeatother

\begin{document}

\singlespace

\begin{flushright} BRX TH-6640 \\
CALT-TH 2018-039
\end{flushright}

\vspace*{.3in}

\begin{center}

{\Large\bf Feynman --- Schwinger Duality  }

{\large S.\ Deser}

{\it 
Walter Burke Institute for Theoretical Physics, \\
California Institute of Technology, Pasadena, CA 91125; \\
Physics Department,  Brandeis University, Waltham, MA 02454 \\
{\tt deser@brandeis.edu}
}
\end{center}

\begin{abstract}
Brief recollections of two giants of physics, their origins, similarities and differences.\end{abstract}

The year 2018 marks the Centennial of my two subjects' birth, duly celebrated by their respective followers. The only missing ingredient is a memorial to their comparative trajectories. That is too tall an order for any one person to fill, but the following may provide some rudiments of a personal assessment.
Let me begin by stating my credentials---and implicit biases. Julian Schwinger (JS) was my thesis adviser, mentor and then lifelong friend, from 1949 when I entered graduate school until his death in 1994. Feynman (RPF) I met early in 1957, when he destroyed my very first grown-up lecture at an international conference, in Chapel Hill, ironically over misuse of path integrals. We met (infrequently) over the years, from Warsaw to Shelter island, until his all-too-early end, in 1988. Like everyone in my field---fundamental theoretical physics---I would read their works as they appeared. One factoid: at some point, I was asked to referee one of their respective late solo papers---alas not their best: Referees are still needed.

What is so striking to any physicist is the polar difference in their personalities and its reflection in their physics, given the close resemblances in their origins, geographically separated only by opposite ends of the subway line (in those days, only the Long Island Railroad): JS in Manhattan, RPF in outer Far Rockaway, the eastern end of New York City. 
Both came from assimilated middle class Jewish families. Yet RPF reveled in his Noo Yawk accent, while JS spoke in mellifluous, cultivated tones, much as Murray Gell-Mann, the next generation Manhattanite (and RPF's office mate), was to do. JS was a manifest wunderkind from the start, attending a (now defunct) super- elite, but public, High School (Townsend Harris), RPF went to plain old Far Rockaway High --- but all public schools were of high level in those Depression days, when teachers had PhDs: there were no College jobs. RPF must have been a wunderkind too, but his schooling was chronologically normal, whereas JS published two papers at seventeen, and seven at nineteen, all highly non-trivial and frontline. Brash and retiringly shy were their obvious respective adjectives. They were both well aware of their talents, however; how could they not? 
Their decade-older predecessor, Lev Davidovich Landau, maintained a logarithmic chart recording the standings of the great. The highest ranking, 0, was assigned to Isaac Newton, Albert Einstein was 0.5, the founding fathers of Quantum Mechanic (QM), Niels Bohr, Werner Heisenberg, Paul Dirac and Erwin Schr\"odinger, all got a one. Landau ranked himself as a 2.5, later self-promoted to a 2, a rating our heroes easily deserve as well; If pressed, I might give JS a 2- and RPF a 2+.

They both had a decisive push when they needed one: RPF acknowledges Mr Bader, his science teacher who got him into MIT, whence he graduated at 21, in 1939. [I cannot resist this example of unintended consequences: Another Far Rock graduate, Martin Annis, went on to found one of the first high-tech firms. I once asked him whether Mr Bader was his hero too---``Don't talk to me about that so-and-so; when I tried to go to MIT, he told me that he would not recommend me because I was nowhere near as smart as the last kid he sent there!  Luckily I made it anyway." I gather that this also happened to other unfortunate successors, including Dyson's at his Public school.] JS's scholastic savior was Lloyd Motz (a nice man, whom I knew), then at City College of New York (another public institution). Apparently JS was too busy with physics to bother with mundane courses there, and was on the verge of expulsion. Motz brought him to see I.I. Rabi, the great physicist at Columbia, across town. He sat on a bench while Motz and Rabi got lost in some arcane QM argument; finally, he could stand it no longer and uttered the word (``Completeness" as I recall) that explained it all---thereby unlocking Columbia. Still, their college, and to some extent their graduate, careers were similar in being primarily self-taught. One of my longer encounters with RPF occurred on the way to Shelter island II in 1983, a small conference celebrating the historic Shelter Island of 1947 where RPF and JS were the stars of the new QED revolution, for which it was convened. To reach the island, located off the very eastern end of Long island, involves a long ferry ride from New London; RPF was on board and decided to corner me, of all other physicists there, to recount his life and times at MIT; being a fast talker, he succeeded in giving me a clear picture of how he and a friend taught themselves QM and other aspects of modern physics. Even at Princeton, his PhD thesis with John Wheeler (himself no slouch), was in good part RPF's alone, and laid the foundations of his postwar work. This brings up the question of mentorship---most of us were helped at the start by one or more professors; that seems not to have been the case here: RPF was not really so indebted to Wheeler and JS's professor---was briefly---Oppenheimer; there is no trace that this had a lasting effect, though the first higher spin (3/2) study did emerge there. This independence seems to be a characteristic of low Landau numbers---they emerge fully formed from Zeus's head!

Wartime came as their research was beginning to take off, and they both plunged into---characteristically 
different---projects: RPF at Los Alamos, where his many exploits are well-known, JS at the MIT Rad Lab,
working on microwaves and other electromagnetic projects; he too was the object of many legends, including his working day that began at 7 PM, as well as his amazing facility with Green functions to solve many--- until then refractory--- problems; it was a tool he never abandoned.
After the war, they were both in immediate demand by Academia, especially since it was expanding rapidly. From a relatively short stint at Cornell, where he was brought by Hans Bethe, RPF moved to Caltech on the West Coast. Instead, JS went down the street to Harvard, where he stayed until---ironically---he too succumbed to California's lure (and Harvard's perceived  disrespect for his physics views), moving to UCLA, across town from Caltech much later. The postwar period was their apotheosis.

That era of QED---quantum electrodynamics is, just as it sounds, the study of the full effect of QM on the coupling of charged particles---electrons, muons, protons---to the electromagnetic field, responsible for almost all atomic (rather than nuclear) phenomena. It had been worked on intensely ever since QM itself started, in the late twenties, but the technical, conceptual and computational difficulties of the next level---the so-called loop effects--- seemed insuperable, particularly the infinities that plagued almost any calculation. Their work has been all too amply parsed at all levels for me to add any commentary. Suffice it to say that there was a good deal of competitive leapfrogging, but at no time any animosity, despite some later attempts by others at finding them, just for dramatic effect. Whenever I saw them together, it was clear that they respected each other, a feeling attested to by their contemporaries. [It is true that JS  barely hid his irritation when a student would lapse into a diagrammatic aside, but it never led to any reprisals.] There was enough glory to go around! In constructing so mighty an edifice---whose  predictions are now known to be accurate to an an unprecedented, almost inconceivable, accuracy, there were of course many other architects, and some inevitable errors were made. RPF's (in)famous ``footnote 13" in one of his key papers about (mis)calculating the Lamb shift or JS's first computation of his famous $\alpha/2\pi$ correction to the electron's magnetic moment---more complicated (it needs no renormalization) than necessary, but these are bagatelles. Their Nobel prizes came a mere two decades later, in 1965.

After reaching this summit, to match it becomes a burden. I believe in the Moses effect: great pioneers are fated never to reach the promised land.  Both did of course do much, characteristically different, fruitful work thereafter: RPF  characteristically went off in many directions, ranging from the pivotal V-A notion (with Gell-Mann, and separately by Marshak and Sudarshan), to partons, as well as novel condensed matter and quantum computing ideas. JS provided the gauge field basis for the future weak interaction theory, proposed the two-neutrino hypothesis, did modern condensed matter pioneering, and discovered a host of other basic results in field theory (his name is attached to many) used to this very day. He then left the mainstream (not a pejorative) to invent Source Theory (``if you can't join 'em, beat 'em") and even dabbled in cold fusion. Both mostly published solo after their early years.

Perhaps the greatest contrast lies in their teaching and mentoring attitudes and their outcomes, where personality is most clearly involved---totally antipodal again: RPF had about $15$ solo graduate students (he quite gruffly turned away Caltech's greatest PhD---Ken Wilson).  Although he dutifully taught graduate courses, he is famous for his Feynman Lectures for Physics freshmen (well, for seniors really!), which I call the undergraduate theoretical minimum, in homage to the great Landau and Lifshits ``theoretical minimum" graduate texts. JS's PhD students number well into the eighties (probably a record of some sort); he taught graduate courses exclusively. Oddly, he later produced some significant popularizations for the BBC. So in teaching they were similar: giving freely of their time, but each in his own way.

These are but a few, if major, aspects of their productive lives. Both had an influence far beyond their teaching and immediate circles---they had leading roles for a long time in defining their field's direction.  
They of course had similarities: not wasting time on the (dare I say) sillier academic obligations self-imposed by  Academia---like most great scientists. I know that JS gave generously, and lent his name, to good causes, even bequeathing his estate to a Foundation. RPF's famous NASA O-Ring service was certainly a time-consuming good deed. I could go on, but my aim has been to use some biographical-scientific aspects of these immortals' lives to show diversity in similarity (or vice-versa). I have consciously stayed away from their (very different) personal lives; this is not that sort of account. I only note that RPF and JS both died of (different) terrible cancers. They --- and their work --- will long be remembered and built upon, even as specifics (yes, even Feynman diagrams) fade. Piano or bongo, vive la difference!

\subsubsection*{Acknowledgements}
I thank A.\ Zee for stimulating the emission of this essay, as well as for many wise suggestions and Thomas Curtright for providing some Feynmania.
This work was supported by grant DOE\#desc0011632.

\end{document}